\documentclass[letter, traditabstract, longauth]{aa} 

\usepackage{graphicx}
\usepackage{natbib}
\usepackage{txfonts}

\begin{document}

\title{{\it Herschel}\thanks{{\it Herschel} is an ESA space observatory with science instruments provided by European-led Principal Investigator consortia and with important participation from NASA.}-SPIRE spectroscopy of G29.96-0.02: fitting the full SED}

\author{ 
	J.M. Kirk\inst{1} \and	
	E. Polehampton$^{2\and3}$ \and
	L. D. Anderson\inst{4} \and
	J.-P. Baluteau\inst{4} \and
	S. Bontemps\inst{5} \and 
	C. Joblin$^{9\and10}$ \and
	S. C. Jones\inst{3} \and
	D.A. Naylor\inst{3} \and 
	D. Ward-Thompson\inst{1} \and
	G. J. White$^{2\and17}$ \and
	A. Abergel\inst{6} \and 
	P. Ade\inst{1} \and
	P. Andr\'e\inst{7} \and 
	H. Arab\inst{6} \and
	J.-P. Bernard$^{9\and10}$ \and
	K. Blagrave\inst{15} \and
	F. Boulanger\inst{6} \and
	M. Cohen\inst{11} \and 
	M. Compiegne\inst{15} \and
	P. Cox\inst{12} \and 
	E. Dartois\inst{6} \and
	G. Davis\inst{13} \and 
	R. Emery\inst{2} \and 
	T. Fulton\inst{19} \and 
	C. Gry\inst{4} \and
	E. Habart\inst{6} \and
	M. Huang\inst{14} \and	
	G. Lagache\inst{6} \and
	T. Lim\inst{2} \and
	S. Madden\inst{7} \and
	G. Makiwa\inst{3} \and
	P. Martin\inst{15} \and 
	M.-A. Miville-Desch\^enes\inst{6} \and 
	S. Molinari\inst{16} \and 
	H. Moseley\inst{18} \and
	F. Motte\inst{7} \and
	K. Okumura\inst{7} \and
	D. Pinheiro Gocalvez\inst{15} \and
	J. A. Rod\'on\inst{4} \and
	D. Russeil\inst{4} \and
	P. Saraceno\inst{16} \and 		
	S. Sidher\inst{2} \and
	L. Spencer\inst{3} \and 
	B. Swinyard\inst{2} \and
	A. Zavagno\inst{4}	
}

\institute{
	School of Physics and Astronomy, Cardiff University, Queens Buildings, The Parade, Cardiff, CF24 3AA, United Kingdom \and
	Space Science \& Technology Department, Rutherford Appleton Laboratory, Chilton, Didcot, Oxfordshire OX11 0QX \and
	Institute for Space Imaging Science, University of Lethbridge, 4401 University Dr., Lethbridge, AB, Canada \and
	Laboratoire d'Astrophysique de Marseille, UMR\,6110 CNRS, 38 rue F. Joliot-Curie, F-13388 Marseille France \and
	CNRS/INSU, Laboratoire d'Astrophysique de Bordeaux, UMR 5804, BP 89, 33271 Floirac cedex, France \and
	Institut d'Astrophysique Spatiale, UMR\,8617, CNRS/Universit\'e Paris-Sud 11, 91405 Orsay, France  \and
	CEA, Laboratoire AIM, Irfu/SAp, Orme des Merisiers, F-91191 Gif-sur-Yvette, France \and
	IPAC, California Institute for Technology, Pasadena, USA \and
	Universit\'e de Toulouse, UPS, CESR, 9 avenue du colonel Roche, F-31028 Toulouse cedex 4, France \and
	CNRS ; UMR5187 ; F-31028 Toulouse, France \and
	University of California, Radio Astronomy Laboratory, Berkeley, 601 Campbell Hall, US Berkeley CA 94720-3411, USA \and
	Institut de Radioastronomie Millim\'etrique (IRAM), 300 rue de la Piscine, F-38406 Saint Martin d'H\`eres, France \and
	Joint Astronomy Centre, University Park, Hilo \and
	National Astronomical Observatories (China) \and
	Canadian Institute for Theoretical Astrophysics, Toronto, Ontario, M5S 3H8, Canada \and
	Istituto di Fisica dello Spazio Interplanetario, INAF, Via del Fosso del Cavaliere 100, I-00133 Roma, Italy  \and
	Department of Physics \& Astronomy, The Open University, Milton Keynes MK7 6AA, UK \and
	NASA - Goddard SFC, USA \and
	Blue Sky Spectroscopy Inc, Lethbridge, Canada 
}

\date{Received 31 March 2010 / Accepted 11 May 2010}

\abstract{We use the SPIRE Fourier-Transform Spectrometer (FTS) on-board the ESA {\it Herschel} Space Telescope to analyse the submillimetre spectrum of the Ultra-compact H{\small II} region G29.96-0.02. Spectral lines from species including $^{13}$CO, CO, [CI], and [NII] are detected. A sparse map of the [NII] emission shows at least one other HII region neighbouring the clump containing the UCHII. The FTS spectra are combined with ISO SWS and LWS spectra and fluxes from the literature to present a detailed spectrum of the source spanning three orders of magnitude in wavelength. The quality of the spectrum longwards of 100\,$\mu$m allows us to fit a single temperature greybody with temperature $80.3\pm0.6$\,K and dust emissivity index $1.73\pm0.02$, an accuracy rarely obtained with previous instruments. We estimate a mass of 1500~M$_\odot$ for the clump containing the H{\small II} region. The clump's bolometeric luminosity of $4\times10^{6}$\,L$_\odot$ is comparable to, or slightly greater than, the known O-star powering the UCH{\small II} region. }


\maketitle

\section{Introduction}

Ultra-compact H{\small II} (UCH{\small II}) regions are small nebulae that surround massive, young stars that are still embedded within a natal cloud \citep{2002ARA&A..40...27C, 2007prpl.conf..181H}. UCH{\small II} regions represent an important evolutionary stage in the formation of massive stars. They may correspond to the still embedded accretion phase while the star is already ionising its surroundings. Their study is critical to our understanding of how stars manage to reach high masses despite the high pressure generated by their growing HII regions. UCHII regions are amongst the brightest Galactic objects at submillimetre wavelengths and can be used as a probe of star formation in distant galaxies. Thus, they make excellent candidates with which to probe all phases of massive star formation and with which to test new instruments.

\object{G29.96-0.02} is an UCH{\small II} region \citep{1989ApJS...69..831W,2002A&A...381..571P} located at a distance of $8.9^{+0.6}_{-0.09}$\,kpc \citep{2004ApJS..154..553S}. The bright IR source at the centre of G29.96-0.02 is an O5-6 star \citep{1997ApJ...490L.165W, 2003A&A...405..175M}, with a luminosity of $L_{bol} \sim 3-4  \times10^{6}$\,L$_{\odot}$ for the \citet{1997ApJ...490L.165W} temperature limits at the \citet{2004ApJS..154..553S} distance. However, models indicate that a cluster of young stars must be present in G29.96-0.02 to account for its observed luminosity  \citep{2003MNRAS.340..799L}. 

At the head of the cometary shaped radio emission \citep{1989ApJS...69..831W} is a small cluster of hot molecular cores ($\sim$300\,K) with masses of 3-11\,M$_{\odot}$ \citep{2007A&A...468.1045B}. These arcsec-sized structures are embedded within an arcmin-sized submillimetre clump called \object{G29.956-0.017SMM} \citep{2006A&A...453.1003T}. \citet{1995ApJ...453..308F} suggested that the cometary shape of the UCH{\small II} region is caused by its expansion westwards into the denser part of the surrounding clump. This clump is part of a larger complex that has been mapped by {\it Herschel}-SPIRE/PACS as part of the Hi-GAL key programme \citep[][Beltran et al. in prep.]{2010molinari}.

In this paper we present new submillimetre spectra taken towards \object{G29.956-0.017SMM} (hereafter G29) with the ESA {\it Herschel} Space Observatory \citep{2010herschel} as part of the ``Evolution of Interstellar Dust'' key programme \citep{2010sag4}. The data were taken with the SPIRE Fourier-Transform Spectrometer (FTS). The SPIRE instrument, its in-orbit performance, and its scientific capabilities are described by \citet{2010spire}, and the SPIRE astronomical calibration methods and accuracy are outlined by \citet{2010spirecal}.

\section{Observations}
\label{obs}

\begin{figure}
\centering{
	\includegraphics[width=0.90\columnwidth]{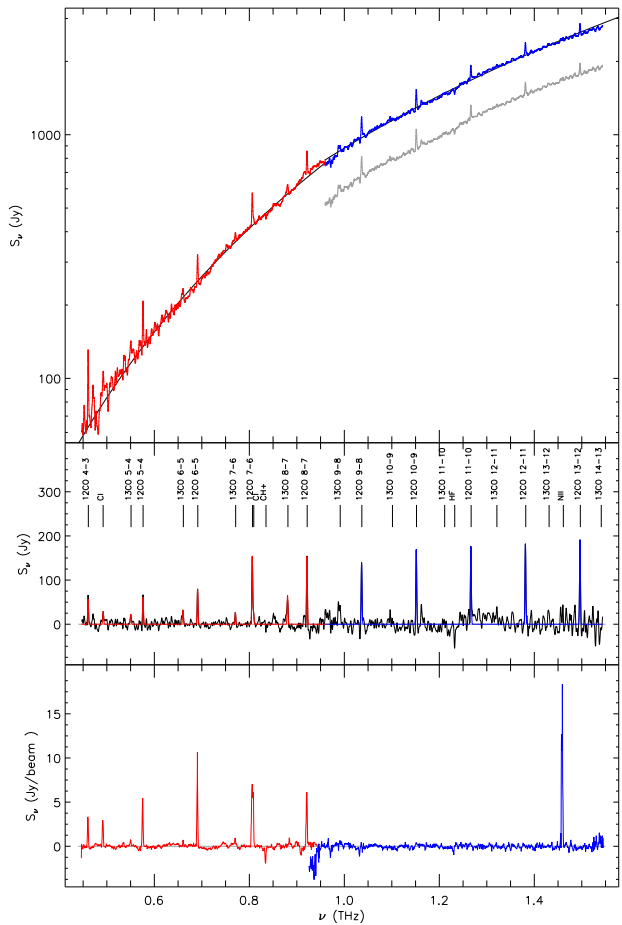}
}
\caption{\label{fig1} SPIRE FTS SLWC3 (red) and SSWD4 (blue/grey) spectra of G29. \textbf{Upper:} On-source spectrum with continuum, point-source calibration. \textbf{Middle:} As above with continuum subtracted. Major lines are annotated. \textbf{Lower:} Mean off-source spectrum, extended calibration. See text for details.}
\end{figure}

G29 was observed with the high-resolution mode of the SPIRE FTS on 13 September, 2009 at 19:49 ({\it Herschel} observation ID, 1342183824). Two scan repetitions were observed giving an on-source integration time of 266.4 seconds. The pointing centre was at a Right Ascension and Declination (J2000) of 18$^{h}$46$^{m}$04.07$^{s}$, $-$02\degr 39\arcmin 21\farcs88 and Galactic coordinates of 29.9561, $-$0.01726. The unapodized spectral resolution was 1.2\,GHz (0.04\,cm$^{-1}$), after apodization \citep[using extended Norton-Beer function 1.5, ][]{2007naylor} this became 2.17\,GHz.

The SPIRE FTS measures the Fourier transform of the source spectrum across short (SSW, 194-313\,$\mu$m) and long (SLW, 303-671\,$\mu$m) wavelength bands simultaneously. Each waveband is imaged with a hexagonal bolometer array with pixel spacing of approximately twice the beam-width. The FWHM beam-widths of the SSW and SLW arrays vary between 17-21\arcsec\ and 29-42\arcsec\ respectively. The source spectrum, including the continuum, is restored by taking the inverse transform of the observed interferogram. For details of the SPIRE FTS and its calibration see \citet{2010spire} and \citet{2010spirecal}.

\section{Results}
\label{res}

\begin{table}
\caption{\label{tab1} Best Fit Line Parameters}
\centering{
\begin{tabular}{lll r@{.}l c r@{ $\pm$ }l }
\hline

Species & Line & Band &  \multicolumn{2}{c}{$\nu$} & $\nu_{offset}$ & \multicolumn{2}{c}{$S_{peak}$} \\

	&	  & & \multicolumn{2}{c}{[GHz]} & [GHz] & \multicolumn{2}{c}{[Jy]} \\
		\hline

12CO & 4-3 & SLW & 461&0 &  0.48 &  57.1 &  12.6  \\
	& 5-4 & SLW &  576&3 &  0.09 & 61.6  &  5.8 \\ 
	& 6-5 & SLW &  691&5 &  0.36 &  80.7 &  1.1 \\
	& 7-6 & SLW &  806&7 &  0.46 &  154 &  2  \\
	& 8-7 & SLW &  921&8 &  0.45 &  154 &  4  \\ 
	& 9-8   & SSW & 1037& &  0.51 &  140 &  6  \\
	& 10-9  & SSW & 1152& &  0.46 &  170 &  7  \\
	& 11-10 & SSW & 1267& &  0.53 &  179 &  15 \\
	& 12-11 & SSW & 1382& &  0.70 &  183 &  7  \\
	& 13-12 & SSW & 1497& &  0.71 &  192 & 14  \\

13CO & 5-4 & SLW & 550&9 &   0.52 &  22.8 &  3.4  \\
	& 6-5 & SLW & 661&1 &   0.74 &  32.9 &  3.0 \\
	& 7-6 & SLW & 771&2 &   0.61 &  25.9 &  4.6 \\
	& 8-7 & SLW & 881&3 &   0.82 &  65.7 &  3.8 \\
	
 {[}C~{\sc i}]  & & SLW &  492&2 &   -0.08 &  29.9 &  6.5 \\
 {[}C~{\sc i}]  & & SLW &  809&3 &   0.50 &  26.8 &  3.3 \\
		\hline
	\end{tabular}
}
\end{table}

The upper panel of Fig. \ref{fig1} shows the apodized SLW (red) and SSW (grey) spectra. Only the data from the central bolometer (C3 for the SLW and D4 for the SSW) as calibrated for a point source are shown for each spectrum. The spectra are dominated by the thermal continuum. Superimposed on this are a series of bright lines, the most noticeable of which are the ladder of CO lines. The shape of the continuum was estimated by masking the CO lines and performing a linear regression of the form $\log S_{\nu} = C + p\log\nu$ to both spectra. The data were best fit by power-laws with indices of $p_{SSW}=2.71$ and $p_{SLW}=3.40$ respectively. There is a disconnect between the two spectra. The offset in power-law constants was $C_{SSW}-C_{SLW}=0.164$, equivalent to a linear factor of 1.45. The SSW spectrum was shifted upwards by this margin and replotted as the blue spectrum under the assumption that the discontinuity was due to differing structure within the SLW and SSW beams. The power-laws for each spectrum are plotted as black lines.

The Levenberg-Marquardt least-squares minimisation package MPFIT \citep{2009ASPC..411..251M} was used to simultaneously fit a catalogue of lines and an 8th-order polynomial to each of the SLW and SSW spectra. It was assumed that the lines were Gaussian and that the linewidths were not resolved. The middle panel of Fig. \ref{fig1} shows the spectra after the polynomial background has been subtracted. The best-fit line shapes are shown in red and blue for the SLW and SSW bands respectively. These show the same ladder of CO lines as the upper panel as well as their $^{13}$CO counterparts. The position of the $^{12}$CO and $^{13}$CO lines are annotated \citep{1998JQSRT..60..883P}. The brightest non-CO lines are the 492 and 809\,GHz lines of [C~{\sc i}]. The 809\,GHz [C~{\sc i}] line is blended with the 806\,GHz J=7-6 line of $^{12}$CO. Table \ref{tab1} lists the transitions, reference frequencies, frequency offsets, and peak flux densities for each significant line fit. 

\begin{figure}
\centering{
	\includegraphics[width=0.9\columnwidth]{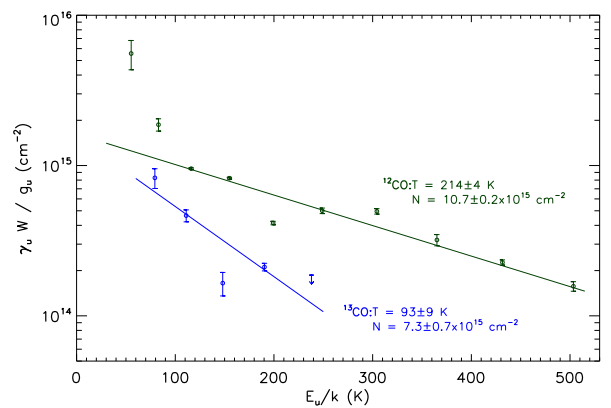}
}
\caption{\label{figRot} Population diagram for $^{12}$CO and $^{13}$CO towards G29. }
\end{figure}

Figure \ref{figRot} shows the rotational population diagram for $^{12}$CO in green and $^{13}$CO in blue. Following \citet{1999ApJ...517..209G} we fit a rotational temperature to each species. The weighted linear fits and results are plotted for each species. The quoted errors are the errors on the fit and do not include uncertainties in beam size, calibration, and sub-structure within the beam. The $^{13}$CO lines come from a greater depth of material than the $^{12}$CO lines and will be more indicative of the temperature of the interior of the clump. The 7-6 transition of $^{13}$CO and the 9-8 transition of $^{12}$CO appear to be lower than the trends. This could be due to the structure in the beam changing between transitions (the beam FWHM changes by a factor of 2.5 between the CO 4-3 and 13-12 transitions).   

\begin{figure}
\centering{
	\includegraphics[width=\columnwidth]{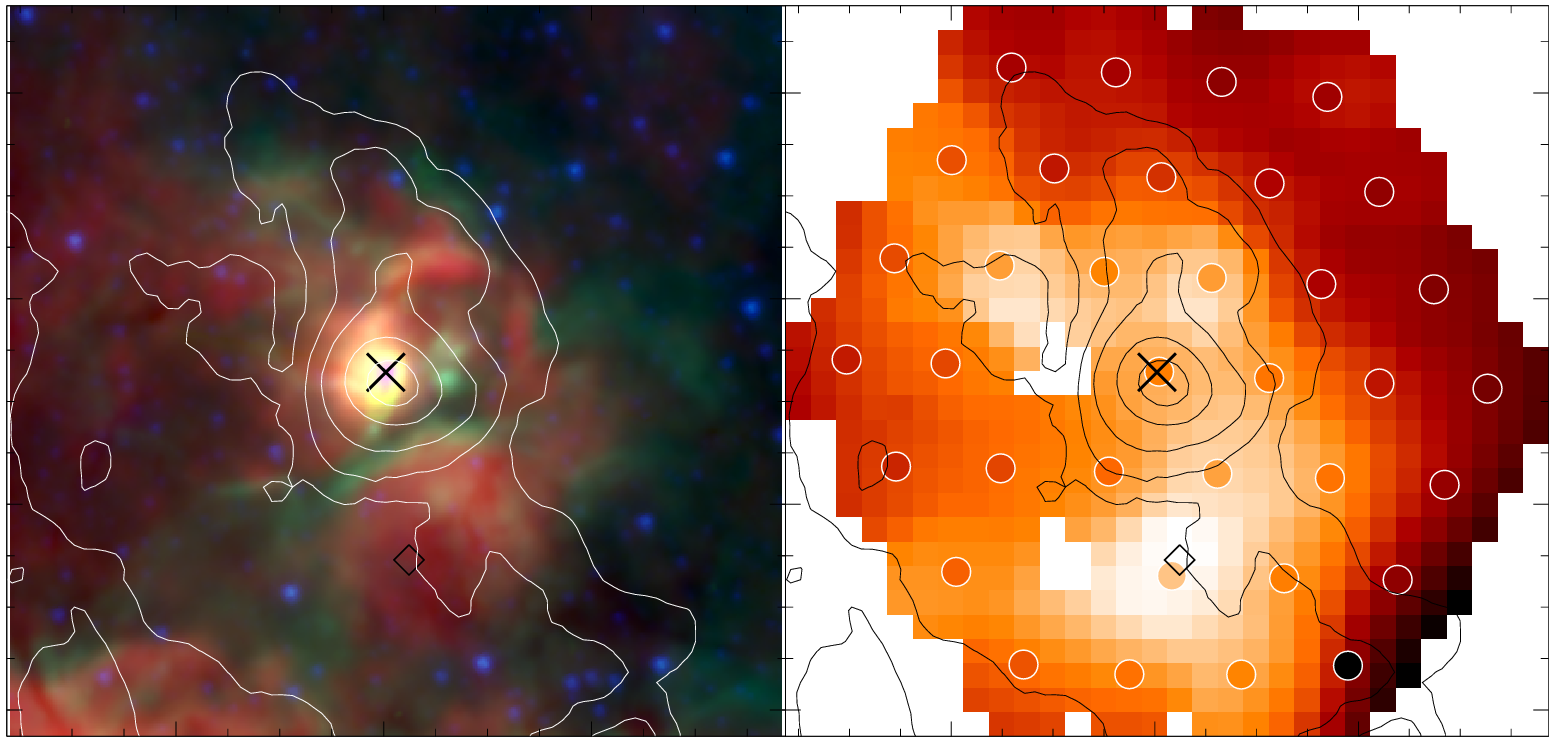}
}
\caption{\label{fig2} {\bf Left:} False-colour map of G29 showing GLIMPSE 4.5 (blue) and 8-$\mu$m (green) \citep{2003PASP..115..953B} and MAGPIS 20cm (red) \citep{2006AJ....131.2525H}. {\bf Right:} [N~{\sc ii}] line flux towards G29. The circles show the position of the bolometers on the sky. The space between them has been interpolated using inverse distance weighting. The X marks the position of the G29.96-0.02 radio source \citep[ 18$^{h}$46$^{m}$03.96$^{s}$, $-$02\degr 39\arcmin 21\farcs5 ]{1989ApJS...69..831W}. The diamond marks the position of the NII peak. The contours are SCUBA 850\,$\mu$m emission (5, 10, 20, 50 and 100 times the local off-source rms) showing the extent of the G29.956-0.017SMM clump.}
\end{figure}

The region around G29 is shown in the left panel of Fig. \ref{fig2} as a false-colour image - blue and green are GLIMPSE 4.5 and 8-$\mu$m \citep{2003PASP..115..953B} and red is MAGPIS 20cm \citep{2006AJ....131.2525H} - and archival SCUBA 850\,$\mu$m contours. The X marks the position of the VLA radio emission associated with the UCH{\small II} region \citep{1989ApJS...69..831W}. The spectra measured towards positions surrounding G29 show several notable differences to the on-source spectra. The lower panel of Fig. \ref{fig1} shows the mean background-subtracted spectrum for co-aligned off-source bolometers. The 835\,GHz line of CH$^+$ is seen in absorption in the SLW band and appears to be probing the ISM \citep[see ][ for a detailed decomposition of the CH$^+$ line for this source]{2010naylor}. The deepest absorption feature is coincident with the 1.232\,THz line of HF \citep{2010neufeld}. The higher-order $^{12}$CO and $^{13}$CO lines are not detected in the SSW band, but the 1.46\,THz fine structure line of [N~{\sc ii}] is strongly detected. This line is present in the extended-source calibration, but not in the on-source point-source calibration. 

A fit to the [N~{\sc ii}] line was performed for all bolometers of the SLW array. The results are plotted on the right panel of Fig. \ref{fig2}. The coloured circle markers show the bolometer positions and the measured line intensities at those positions. For these early results the bolometer array was not offset to create a fully sampled map. To compensate for this the pixels between the bolometers have been interpolated using an inverse distance weighting (modified Shepard's) method. The southern [N~{\sc ii}] peak is coincident with a region of 20cm emission that is bounded to the north by a 8-$\mu$m filament suggesting that it is a separate HII region to the G29 UCHII region. Its diameter is $\sim$1\arcmin\ (2.5\,pc at 8.9\,kpc). The morphology around G29 is clearly complex and a more detailed, fully-sampled study will be required to disentangle the various components. As stated, this preliminary map was not fully sampled, but it does shows the potential of the FTS as a mapping spectrometer.

\section{Spectral energy distribution}
\label{sed}

\begin{figure*}
\centering{
	\includegraphics[width=0.77\textwidth]{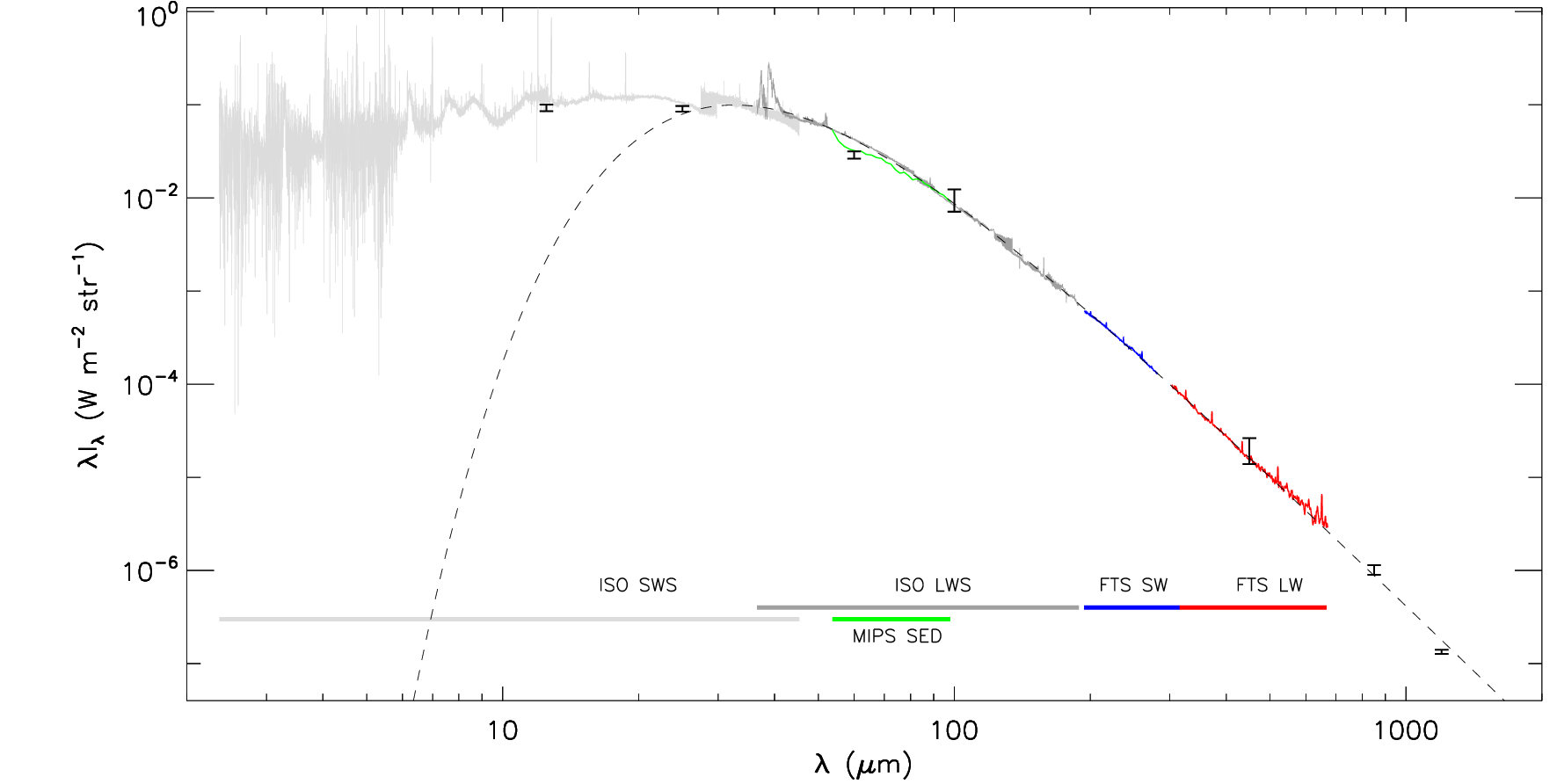}
}
\caption{\label{fig3} The G29 spectral energy distribution. SPIRE SWS-D4/SLW-C3 spectra are shown by the blue/red lines. 
The light and dark grey spectra are archival data from the ISO Short Wavelength Spectrograph \citep[SWS]{1996A&A...315L..49D} and ISO Long Wavelength Spectrograph \citep[LWS]{1996A&A...315L..38C} respectively and were originally published by \citet{2002A&A...381..571P}. Archival MIPS SED mode data \citep{2004ApJS..154...25R} are shown by the green line.
The error bars show IRAS 12.5-100\,$\mu$m \citep[IRAS 18434-0242 is coincident with G29]{1988iras....7.....H}, SCUBA 450 and 850\,$\mu$m \citep{2006A&A...453.1003T}, and IRTF 1.3\,mm \citep{1986A&A...154L...8C} data points. The FTS SED data have been shifted to the same calibration as the ISO LWS data. A best fit SED is shown by the dashed line.
}
\end{figure*}

Figure \ref{fig3} shows the spectral energy distribution (SED) measured towards G29. SPIRE FTS SLW-C3 and SSW-D4 spectra are shown in the same colours (red and blue) as Fig. \ref{fig1}. A search was made for archival data coincident with G29. For consistency with the FTS data the downloaded spectra and data points were converted into intensity units by dividing by their wavelength-dependent beam area or, where relevant, their aperture area. The archival data is plotted in Fig. \ref{fig3} and described in the Figure caption. 

Figure \ref{fig3} does not include the relative correction applied in Fig. \ref{fig1}. It was found that the FTS SLW data were marginally higher that the SCUBA 450$\mu$m data point and that the first point of the FTS SSW spectrum was significantly higher than the ISO Longwave Spectrograph (LWS) final data point. It was assumed that the three sections of the long wavelength SED (ISO LWS, FTS SSW, FTS SLW) followed the same single temperature greybody and that the offsets between them were due to differences in their absolute calibration. A greybody equation was fitted to these three spectra and offsets between them accounted for by giving the flux from each section a multiplier coefficient. The greybody had the form
\begin{equation}
I_{\nu} = C_{X} B_{\nu}(T) \left(1- e^{ -(\nu/\nu_{c})^{\beta}}\right)
\end{equation}
where $C_{x}$ is the coefficient, $B_{\nu}(T)$ is the Planck function at temperature $T$, $\nu_c$ is the frequency where the emission becomes optically thin, and $\beta$ is the dust emissivity index. The data were resampled into evenly spaced bins in log-wavelength space to prevent the differing density of data points between the ISO LWS and the SPIRE FTS corrupting the minimisation. 

The best-fit to the SED was $T=80.3\pm0.6$\,K, $\beta=1.73\pm0.02$, and $\nu_{c}=20.0\pm0.5$\,THz (equivalent to $\lambda=15.0\pm0.4$\,$\mu$m). We note that the SED temperature is similar to the $^{13}$CO rotational temperature from Fig. \ref{figRot}. The multiplicative coefficients for the FTS SSW and FTS SLW were found to be $9.3\pm0.08$ and $4.04\pm0.06$ respectively, the ISO LWS coefficient was fixed at 1.0. The SPIRE FTS spectra have been aligned to the ISO LWS calibration. A single temperature greybody well fits the data longwards of $\sim40$\,$\mu$m. However, fitting a single component greybody to a complicated source like G29 can only yield a broadly characteristic temperature. The relatively flat SED shortwards of the peak shows that there must be several hotter temperature components. Nevertheless, Fig. \ref{fig3} does show that the long wavelength the slope of the SPIRE FTS data is consistent with the slope of the literature data.   

In the following paragraphs we adopt the distance of 8.9\,kpc \citep{2004ApJS..154..553S} for G29 and assume that the majority of the emission comes from a region comparable to, or smaller than, the Hershel beam at 250\,$\mu$m ($\theta$=18\arcsec) -- (at 850\,$\mu$m, 60\% of the extended emission towards G29 is within the central SCUBA 14.7\arcsec\ beam, \citep{2006A&A...453.1003T}). Based on the fitted dust temperature, G29's published 850\,$\mu$m peak flux density \citep{2006A&A...453.1003T}, and making typical assumptions \citep[e.g.][]{2005MNRAS.360.1506K} we estimate the mass of the G29 clump to be $M \sim 1500$\,M$_{\odot}$. As with the fitted temperature, the actual mass will depend on the internal profile/geometry of the clump, but we note that this mass is similar to the median mass (940\,M$_\odot$) of the infra-red dark clouds (IRDCs) from which high-mass stars are believed to form \citep{2006ApJ...641..389R}.
 
The luminosity integrated under the fitted greybody in the range 2-2000\,$\mu$m is $L_{Dust} = 61 D^2 \theta^2$~L$_{\odot}$ where $D$ is the distance to the source in kpc and $\theta$ is the diameter of the emitting area.  The above assumptions give a luminosity of $L_{Dust} = 1.6\times10^6$\,L$_\odot$. This agrees with the luminosity of 10$^{6}$\,L$_{\odot}$ estimated from IRAS measurements alone \citep{1991ApJ...372..199W}. Likewise, the bolometric luminosity in the range 2-2000\,$\mu$m, interpolating to the fitted SED at $\lambda>650$\,$\mu$m, is $L_{bol} = 154 D^2 \theta^2$~L$_{\odot}$. The greybody luminosity is $\sim$40\% of $L_{bol}$. At the assumed distance $L_{bol} = 4.0\times10^6$\,L$_{\odot}$. The $L_{bol}$ of the clump containing the UCH{\small II} should equal the luminosity of the driving sources if the region is in equilibrium.  Our calculated $L_{bol}$ is on the upper limit of the range of luminosities for the identified single O-star in G29 \citep[$3-4\times 10^{6}$\,L$_{\odot}$, ][]{1997ApJ...490L.165W} which may support the idea that the luminosity from more than one star is powering the reprocessed SED \citep{2003MNRAS.340..799L}. 

\section{Conclusions}

We have presented new SPIRE FTS 190-670$\mu$m spectra of the submillimetre clump G29.956-0.017SMM which contains the G29.96-0.02 UCH{\small II} region. The impressive capabilities of the SPIRE FTS have allowed us to simultaneously observe both the dust continuum and prominent spectral lines towards G29. We have conducted basic line-fitting and shown the distribution of [N~{\sc ii}] emission towards G29. While this preliminary map was not fully sampled it does show at least one other HII region neighbouring G29 and demonstrates the amazing potential of the FTS as a mapping spectrometer. We have reconstructed the SED of G29 using the FTS spectra and archival measurements and have shown that the FTS calibration is broadly consistent with earlier observations. The combined data-set allowed us to fit a precise greybody with $T=80$\,K and $\beta=1.73$. Based on a distance of 8.9\,kpc we estimated the mass of G29 to be approximately $1500$\,M$_{\odot}$. The calculated luminosity of the G29 clump is slightly greater than the known O-star at the centre of the UCH{\small II} region. 

\begin{acknowledgements} We thank D. Neufeld for identifying the 1.2\,THz HF absorption feature. JMK acknowledges STFC funding, while this work was carried out, under the auspices of the Cardiff Astronomy Rolling Grant. SPIRE has been developed by a consortium of institutes led by Cardiff University (UK) and including Univ. Lethbridge (Canada); NAOC (China); CEA, LAM (France); IFSI, Univ. Padua (Italy); IAC (Spain); Stockholm Observatory (Sweden); Imperial College London, RAL, UCL-MSSL, UKATC, Univ. Sussex (UK); and Caltech, JPL, NHSC, Univ. Colorado (USA). This development has been supported by national funding agencies: CSA (Canada); NAOC (China); CEA, CNES, CNRS (France); ASI (Italy); MCINN (Spain); Stockholm Observatory (Sweden); STFC (UK); and NASA (USA).
\end{acknowledgements}

\bibliographystyle{aa}
\bibliography{g29}

\end{document}